\documentclass[review]{elsarticle}

\usepackage[draft]{hyperref}
\usepackage{amsmath,amssymb,amsfonts}
\usepackage{algorithmic}
\usepackage{graphicx}
\usepackage{textcomp}
\usepackage[ruled,vlined]{algorithm2e}
\usepackage{setspace}
\usepackage{subfigure}
\usepackage{setspace}
\usepackage{enumerate}
\usepackage{soul}
\usepackage{ccaption}
\usepackage{xcolor}

\journal{Journal of \LaTeX\ Templates}

\begin{document}

\begin{frontmatter}

\title{Cellcounter: a deep learning framework for high-fidelity spatial localization of neurons}

%% Group authors per affiliation:
\author[1]{Tamal Batabyal\corref{mycorrespondingauthor}}
\cortext[mycorrespondingauthor]{Tamal Batabyal}
\ead{tb2ea@virginia.edu}

\author[1]{Aijaz Ahmad Naik \fnref{aan}}
\fntext[aan]{Author left the University of Virginia. He is now at the National Institute of Health, Bethesda, MD 20892, United States.}

\author[3]{Daniel Weller \fnref{dsw}}
\fntext[dsw]{Author left the University of Virginia. He is now at KLA Corporation, Ann Arbor, MI, USA 48105, United States.}

\author[1]{Jaideep Kapur}

\address[1]{Department of Neurology, School of Medicine, University of Virginia,  Charlottesville-22904, Virginia, United States}
\address[3]{Department of Electrical and Computer Engineering, University of Virginia, Charlottesville-22904, Virginia, United States}

\begin{abstract}
Many neuroscientific applications require robust and accurate localization of neurons. It is still an unsolved problem because of the enormous variation in intensity, texture, spatial overlap, morphology and background artifacts. In addition, curation of a large dataset containing complete manual annotation of neurons from high-resolution images to train a classifier requires significant time and effort. We present Cellcounter, a deep learning-based model trained on images containing incompletely-annotated neurons with highly-varied morphology and control images containing artifacts and background structures. Leveraging the striking self-learning ability, Cellcounter gradually labels neurons, obviating the need for time-intensive complete annotation. Cellcounter shows its efficacy over the state of the arts in the accurate localization of neurons while significantly reducing false-positive detection in several protocols.         
\end{abstract}


\end{frontmatter}

%\linenumbers

% ________________________ %
\section{Introduction}
\label{intro}
Neurons are the primary unit for complex computation in our brains. They form extensive and intricate networks to generate a complex behavioral repertoire. Neuroscientists seek to map the whole brain circuits engaged in various behaviors. This effort is spurred by the arrival of genetically-engineered system-probing accessories (activity reporter mice, chemogenetics, optogenetics, etc.), powerful imaging methods (confocal, light-sheet, EM microscopes etc.), and the availability of nearly-unlimited storage, ~\cite{dabrowska2019parallel, moser2008place}. Empirical evidence suggests that such circuits are hierarchical, starting at the cellular resolution~\cite{helmstaedter2013cellular, lichtman2014big} to brain regions~\cite{sporns2013making, rubinov2010complex}. Anatomical mapping has provided useful insights in laying out the holistic association of a function or behavior to neuronal ensembles at specific regions in our brain~\cite{sporns2013making}. 
Spatiotemporal localization of neurons is an indispensable tool for circuit mapping. It is utilized to quantify measurable entities like activation map and neurodegeneration, which are correlates of several problems, such as epilepsy~\cite{kramer2012epilepsy, stam2014modern}, Alzheimer's disease, memory and learning~\cite{josselyn2020memory, roy2019brain}. It requires a concerted manual effort, annotating neurons in cases of \textit{C.elegans} ($302$ neurons) or a mollusk ($20,000$ neurons). However, for a rodent brain with billions of neurons, annotation in 3D demands automation.        

The efficacy of counting and localizing a cell type largely depends on several processing steps, such as variations in protocols, slide preparations, staining, and microscopy. While executing these steps, biological noise, technical noise and artifacts are ubiquitous in data. Apart from noise and artifacts, there are several commonly encountered challenging scenarios, such as (1) different cell subtypes with diverse morphology, (2) non-uniformly scattered and densely packed cells, (3) regions with overlapped cells, and (4) significant variation in pixel intensities. Therefore, the algorithms to extract features, either by explicit customization or implicit transformation, for cell localization, counting, and classification, are highly context-selective, protocol-specific, and problem-related.

In this work, we present a deep learning-based pixel-level regressor, called Cellcounter, consisting of convolutional layers and a non-local attention module (shown in Fig.~\ref{fig1: workflow}) and it is trained on nearly $350,000$ image patches post-augmentation. In practice, a deep network demands a formidable corpus of data during training, and  the data curation process to fully annotate neuron cell bodies in a single image is significantly time-intensive. That because we are interested in neurons in this work, we accordingly ask a relevant question: is it feasible to properly train our network with partially annotated data and retrieve the missing annotations during training by the network itself? To find an answer to that question, \textit{we simultaneously test Cellcounter's self-learning ability on the training dataset to complete the annotation (or equivalently sampling)} and \textit{generalization on the validation and unseen datasets}. As \textit{improved detection on a handful of image patches does not necessarily imply effective generalization}, we provide the detection results on whole slices as well as different regions. 
Based on the quantified results, we find that complete labeling of stained brain slices or image patches is not necessary for localizing neurons. In addition, in contrast to conventional cell detection frameworks, \textit{we introduce control images (artifacts, noise and unwanted structures) and show that it greatly improves the overall performance}.

For people with deep learning or of similar backgrounds, Cellcounter configurations can serve as potential pre-trained models to begin training with customized datasets. Finally, this tool is accessible to investigators who are not familiar with coding and deep learning. Straightforward loading of image files (maximum intensity projected (MIP) images in .tiff/.nd2 file (NIS Elements)/ layer-specific stack of .tiff files) as demonstrated in the flowchart in Fig.~\ref{fig1: workflow} and the selection of staining-specific configuration would produce labeled images as output.   

\noindent\textit{Contribution}: 
\begin{enumerate}
    \item We presented a deep neural network, named Cellcounter, consisting of fully convolutional layers with a non-local attention module between the encoding and decoding stages. The network detects the centers of neuronal cellbodies from images and registers depth (`Z' coordinates) if image stacks are presented. The false positive and negative detection rates are significantly low compared to the state of the arts (Fig. \ref{fig2: Dataa}). 
    
    \item We curated control images, containing several artifacts, clutters, and noise, and used them for training to aid in robust prediction (Fig.~\ref{fig2: Data}). We presented the results on entire slices (section~\ref{ctrldata},Fig.~\ref{fig4: Results}, Supplementary Fig. 1 and 2), considered different staining modalities, investigated the variations in images and neuronal morphology, the sources of artifacts and noise in detail. Unlike many contemporary works, Cellcounter does not need images with high signal-to-noise ratio (Fig.~\ref{fig2: Data} C-a, C-b, B-f; Supplementary Fig. 2). 
    
    \item We showed that image slices do not need to be fully annotated. Leveraging the self-learning ability of Cellcounter (which we validated in section~\ref{self-learning}) and self-similarity of neuronal morphology, partial annotation (Fig.~\ref{fig5: MoreResults} B) is sufficient to train Cellcounter, where Cellcounter completes the annotation process during training by itself. This saves a lot of time for manual annotation.

\end{enumerate}
\section{Related works}
%new add
Cell counting and localization fall under the category of object detection and recognition in the context of machine learning. In literature, it is envisaged as two different problems owing to the fact that cell density can be measured without explicit localization. 
Prior to the deep learning era, the state of the arts extracted customized features which were subsequently channeled to a classifier to discriminate between objects and represent classes. There are several, popular discriminatory features, such as scale-invariant feature transform (SIFT)~\cite{lowe2004distinctive}, histogram of oriented gradients (HOG)~\cite{dalal2005histograms}, speeded up robust features (SURF)~\cite{bay2006surf}, PCA-SIFT~\cite{ke2004pca}, non-redundant local binary pattern (LBP)~\cite{nguyen2010object} and Gabor filter bank~\cite{jain1997object}. In another avenue, several filters were proposed to enhance the `objectness', that was integrated into the hand-designed features. These filters aided not only in object localization, but in delineating the morphology of the objects. For example, Level sets~\cite{sussman1994level}, active contours~\cite{chan2001active,caselles1997geodesic}, and Frangi's vesselness method~\cite{frangi1998multiscale} were employed to quantify the shape of an object post-localization. These methods were proved effective in object classification and localization~\cite{budai2013robust} in limited scenarios. 

With such assortment of hand-crafted features, numerous high-end classifiers and classification strategies were developed and rigorously investigated to maximize accuracy in cell segmentation as well as localization. Authors in ~\cite{tikkanen2015training} proposed a set of possible regions for each non-overlapping cells and the merit of each proposed region was evaluated using SVM. Segmentation based deep learning tools, such as U-Net~\cite{falk2019u}, networks based on U-Net~\cite{guo2019sau}, Cellpose~\cite{stringer2021cellpose} are available for the segmentation of boundaries of cell type with which the model is adequately trained for.
Although they achieved improved performance in cases of several cell types, we have noticed that segmentation based cell detection  perform poorly to detect neurons in scenarios, where multiple neurons overlap in a brain slice or a neuron is partially occluded by axonal fibers. Semi-automated methods, which are cumbersome in practice, using ImageJ or Imaris tools to segment neurons in a single slice are fraught with the same problems.   

To alleviate the severity of the problem, the recent cell detection \emph{protocol} follows that the center pixel of each cell is marked as 1 (`dot') with rest pixels as 0. In contrast to segmentation where a cell is marked with a continuous boundary, such binarization lured the researchers to solve such problems following either unsupervised or supervised approaches. However, the dot-annotated label images are significantly sparse and posed problems during regression. As a result, it is becoming a practice to convolve the dot-annotated images with a standard kernel~\cite{lempitsky2010learning, fiaschi2012learning, xie2018microscopy, paul2017count, walach2016learning} or to use problem-specific coding schemes~\cite{xie2015beyond, liang2019enhanced}. 
Using such transformed labeled images, the output of the regression can be either an estimated count of total number of cells present in an image or a learned density map. Density map provides a flexibility in a sense that given a region selected within the image, the number of cells that exist in that region can be approximated by the summation/integration of density values of the region. As mentioned earlier, this procedure do not need explicit localization.
Using the kernelized label images for training, there are several recent approaches that can be again broadly classified in two categories, which are the methods that use state-of-the-art features and data-driven methods involving deep learning. 

Among the works using state-of-the-art features on kernelized dot-annotated image protocol, noteworthy is~\cite{lempitsky2010learning}, where the authors attempted to formulate a regression framework that provides a density map, a measure of `objectness' of a pixel. Their approach was robust and versatile, and carefully avoided the hard counting (i.e. the way a person counts) at the expense of extremely poor cell localization. Similar footsteps were followed in~\cite{fiaschi2012learning}, where the quality of the output density map was improved but the precision of localization was ignored. By efficiently applying ridge regression to estimate the density map, authors in~\cite{arteta2014interactive} developed an interactive platform where a feature vocabulary was learned on-the-fly as a user tags a neuron with a dot. Here, the process of counting was automatic but the localization was performed manually.

In deep learning approaches, the mapping between the raw image input and the kernelized dot-annotated image as the output is performed implicitly, encoding complex input-output relationships by way of tuning thousands of weights as parameters. Over years, techniques, such as batch normalization, dropout, and pooling provided resilience to the overfitting problem. Based on MatConvNet~\cite{vedaldi2015matconvnet}, Xie \emph{et al}.~\cite{xie2018microscopy} proposed a parallel architecture that compares two fully convolutional regression networks (FCRN) to provide the density map at the output for cell counting and evaluated their network on retinal pigment epithelial (RPE) and precursor T-cell lymphoblastic lymphoma (precursor T-cell LBL) cells. For cells with fairly regular shapes and small structural variation, with spatial characteristics, such as high packing density and non-overlapping cell boundaries, the method can also localize them by finding local maxima in the density map. 
Liu and Yang~\cite{liu2017novel} proposed a cell detection algorithm using CNN and maximum weight independent set (MWIS). Although their approach gained significant boost in the performance, the construction of graph for a large number of cells and thereafter, the execution of MWIS are computationally expensive.

Instead of density map, Cohen \emph{et al.}~\cite{paul2017count} developed a framework for redundant counting based on the receptive fields of a regression network, which is effective in much complex scenarios, such as overlapping cells. However, as mentioned by the authors, the accuracy of localization was sacrificed in addition to a set of other computational problems, including the effect of sparse or no-cell background on the regression scores. Xue \emph{et al.}~\cite{xue2016cell} partitioned an input image into a set of image patches and aggregated the counts per patch to obtain the total count. The work used scalar count approach and provided a spatial density map per image. However, their work suffers from poor localization issue and anomalous annotation of cells at the patch boundaries.  Class-agnostic counting in~\cite{lu2018class} implemented a generic matching network (GMN), relying on the self-similar morphology and/or shape of objects. The detection and localization were significantly improved except having possible caveats, such as confounding objects (ex. densely packed neurons and blood vessels in a brain slice) and wide variation in shapes and morphology~\cite{batabyal2020neuropath2path, batabyal2018elastic}(ex. between cerebellar purkinje cells and hippocampal parvalbumin interneurons). To improve the cell detection, layer-wise gradient boosting and selective sampling strategy were used in \cite{walach2016learning}, where the labeled images are convolved with a Gaussian kernel. Utilizing the U-Net~\cite{ronneberger2015u} architecture, authors in~\cite{guo2019sau} introduced a self-attention module, which is a non-local weighted mean measure, and used online batch-normalization to detect cells.  Over time, it is becoming apparent that developing a versatile deep learning model configuration that, upon rigorous training, can result in the precise detection and localization of various cell types is overly challenging. In this work, we confine ourselves to improving the quality of localization and counting only neurons from confocal microscopy images (see section~\ref{app} for applications).

\begin{figure}[!ht]
\vspace{-.2cm}	
	\centering
	\includegraphics[width=12cm, height=15.5cm]{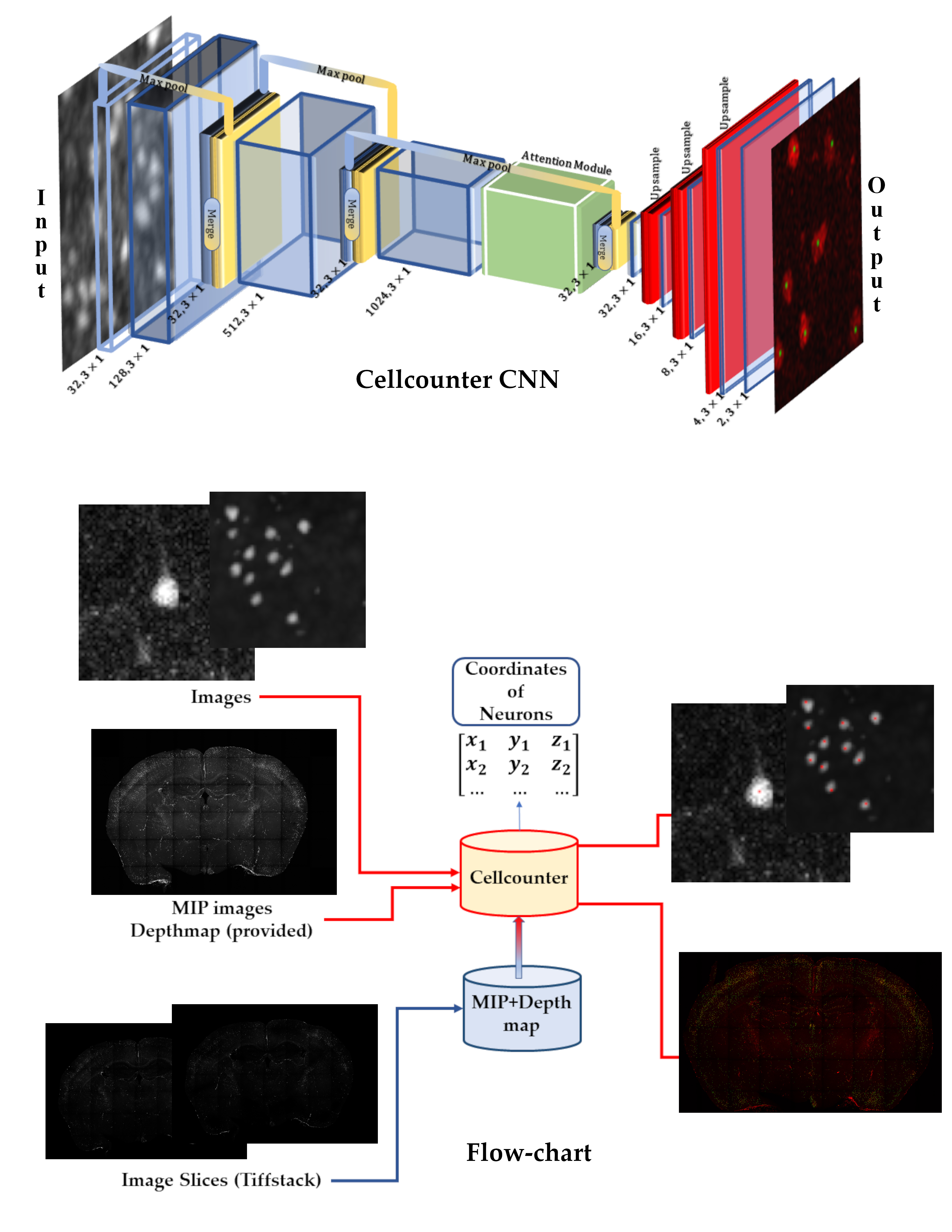}
	\caption{   }
	\label{fig1: workflow}
	\vspace{-.1cm}
\end{figure} 
\begin{figure}[!ht]
  \contcaption{\small (\textbf{Top}) \emph{The configuration of our proposed Cellcounter}. It has $13$ convolutional layers with the number and size of each layer mentioned in the figure. 'Merging' module places two $32$-filter layers back to back, followed by a convolutional operation, such that it carries and passes deep-layer structural details. We added non-local means based attention network~\cite{guo2019sau,wang2018non}. (\textbf{Bottom}) \emph{A schematic showing the work-flow of our model framework}. To localize/label the neurons, images / maximum intensity projected (MIP) image of a stack of images / image stack (.tiff) can be given as input and users need to select the appropriate option when prompted. If `Z' (depth) coordinate is required, users are expected to provide the `Z' map (an image having the same dimension as the input image and registering the slice number where the maximum at each pixel has occurred out of $20$ slices) \textit{when} MIP image is selected as an input. For image stack (for example, a $200\mu$ thick slice containing $20$ slices of $10\mu$ optical thickness), our code automatically computes the MIP image and saves the `Z' map  for later use. Next, Cellcounter labels each neuron with \textit{single pixel} that approximates the center of neuron cell body. 
  Once the cells are labeled, user can select regions by drawing a contour around the region (Not shown here). The number of labeled neurons inside the region, including on the boundary will be displayed.  }% Continued caption
\end{figure}

% new add
\subsection{Challenges}
As previously mentioned that the efficacy of cell counting and localization depends on a variety of preprocessing steps. A cleanly stained image slice may not be obtained at all stages in an experiment. In addition, the quality of slicing, staining and imaging contains human bias. In this section, we describe the problems and challenges regarding detection and localization of neurons in an image slice (mice brain).

\noindent\textit{Downsampling}: Downsampling operation introduces morphological distortion in the neurons. A downsampling rate of $5$ or more thus yields significant distortion in the neuronal morphology. This effect is described in Fig.~\ref{fig2: Data}(D), where two neurons are progressively downsampled. After a downsampling factor of $4$, the arbor of each neuron disappeared  and is visually indistinguishable with a trace of blood vessels (Fig.~\ref{fig2: Data}(E-d)) or localized stain patch (Fig.~\ref{fig2: Data}(A-h)). 
We did not perform downsampling on any image or image patches in our experiments. That because downsampling alters neuronal morphology, precaution (for example re-training Cellcounter with downsampled image patches) should be taken when applying Cellcounter on downsampled images.  \\

\noindent\textit{Clutter/artifacts/Noise}: The maximum intensity projection (MIP) operator introduces unwanted clutters in the form of axonal fibres superimposed on the neurons, thereby occluding the neurons partially and sometimes completely. Axonal fibres, shown in Fig.~\ref{fig2: Data}(A-i,c,e) are present in all the slices at places, such as corpus callosum, fimbria, fornix, and alveus. It is difficult to filter them out. Like axonal fibers, blood vessels are also omnipresent in slices and they share similar intensity profiles as of a neuron post-registration. Axonal fibers and blood vessels are indigenous clutters. Axonal fiber tracts have long spatial span and dimmer intensity profiles. Blood vessels are of shorter span and have bright intensity in each slice. 

There are several extraneous artifacts. Microcracks caused by intracranial electrodes at stereotaxic targets, parallel tears and splintering out in tissues, parallel undulations in brain slices, folds and wrinkels, oblong and irregular holes, blocky artifacts while image stitching are few examples of such artifacts. The neurons that fall along the microcracks or in irregular holes are usually undetected by a CNN. Undulations generally have an adverse impact on the quality of registration, producing undesirable long and bright seams after registration. However, they have negligible effect on the accuracy of neuron detection and localization. 

We carefully curated a large set of image patches (see Fig.\ref{fig2: Data} (A)), which we call the control set and this set represents all the clutter, artifacts and noise. The output label image corresponding to each control image is a null-image (all pixels are zero) and this suppression helps in robust detection of neuronal cellbodies. To obtain better precision, the label image could contain negative values as a penalty. We consider it as our future work. \\

\noindent\textit{Annotation error}:
Owing to the large volume of neurons with wide variation in morphology in an MIP image, it is time-intensive to label all the neurons in the slice. We visually confirmed differences in neuronal morphology in different regions (cortices, hippocampus, thalamus etc.) and non-uniformly sample the neurons . Here, \textit{sampling} refers to the registration of the $(x,y)$ location of a neuron cellbody by clicking on the center pixel inside the cellbody. Due to intensity inhomogeneity of stained images and the existence of overlapped neurons, it is sometimes difficult to find the center of each neuron. This leads to error in manual annotation.  We allow minor deviation of the detected location from the cell centers (see Fig.~\ref{fig4: Results}(G) arrowhead). \\

\noindent\textit{Patch boundary problem}:
While labeling a neuron, we consider single pixel based annotation. The labels of the neurons that are along patch boundaries may fall outside the patch, leading to an anomaly of whether or not those cells are to be tagged by Cellcounter and be considered in accuracy of detection. After labeling each patch by Cellcounter during test, all the patches are assembled to output the original image with labeled neurons. The neurons that fall on the boundaries may be tagged more than once (overcounting). This problem can be eliminated by retaining one of the detected labels and cancelling the rest which are in proximity. However, we observed that this routine affects closely-spaced labels which correspond to a set of overlapped neurons. \\

\noindent\textit{Error in depth (Z coordinate) estimation}:
As explained in Fig.~\ref{fig1: workflow}, that MIP yields the `Z' coordinate of a neuron cellbody. The MIP image is computed on a stack (generally $10$ or $20$; $10\mu m$ or $40\mu m$ in our experiments) of stained image slices. It is the slice number where the center pixel has maximum intensity is recorded for `Z' coordinate. Therefore, the estimation of depth is sensitive to the slice order.  The major bottleneck stems from how the slices are placed in the wells following cryostat/vibratome (slicer). The wells (where slices are kept after cutting) are manually numbered to keep track of the chosen axis. For example, while preparing coronal sections, the wells are manually numbered in the order of depth from the anterior (Nose) to the posterior (Cerebellum) of a mouse brain. Ideally, each slice should be placed in a single well maintaining the order of depth, which is resource-intensive and laborious. This prompts us to place multiple slices ($3-4$) in a well, where the order of depth of slices in each well is compromised and can not be retrieved later. For a $40\mu$ thick slices, there is at most $120\mu -160\mu$ error in thickness with $3-4$ slices in each well. This error can not be rectified by using standard mouse brain atlases, such as Paxinos and Franklin stereotaxic coordinate atlas, where the imaged coronal sections are shown with $300\mu$ distance between consecutive coronal slices. 
We reduce the depth error of a neuron by averaging the `Z' values (in a $3\times 3$ window) around the pixel ($x,y$) that is detected by Cellcounter. \\

\noindent\textit{Post-registration distortion} (Applicable only for registered slices) : 
We performed registration using~\cite{liang2018structure} on few slices to test Cellcounter's ability (Supplementary Fig. 1). 
We observed distortions in registered images in terms of intensity and neuronal morphology. A neuron in an unregistered tdTomato images shows intensity inhomogeneities among the pixels that define the cell body, whereas the same neuron followed by the registration with the fused (Dapi+tdTomato) images exhibits diffused appearance. The intensity enhancement or scaling to tdTomato images prior to fusion partly contributes to the diffused intensity. Moreover, the blurred feature of Dapi staining contributes to the diffused appearance. 
In some of the regions, especially layer $1$ of the ventral retrosplenial, layer $2/3$ of the dorsal retrosplenial and the posterior parietal association, layer $2/3$ and $4$ of the somatosensory cortical areas, the diffusion is noticeable because the diffusion causes the boundaries between some of the neighboring cells disappear. This fact encourages us to use coding strategy as given in~\cite{liang2019enhanced} instead of the Gaussian convolution based labeling strategy for training our CNN. \\

\noindent\textit{Risk of misclassification}  :
Misclassification of cells typically includes cells which are unlabeled and other objects which are falsely labeled as cells. There are statistical measures in terms of Type I and II errors to quantify them. There is an additional problem in the context of misclassification - overcounting a cell (multiple maxima detected on a cell). The individual risks of these errors on the understanding, prognosis and diagnosis of a problem is largely context-specific and sometimes not quantifiable. In the following paragraphs, few relevant contexts are presented.

When considering recruitment of a brain region during seizure propagation, the bounds on the misclassification (Type I and II) are not so tight as it is merely a binary decision based on the density of neuronal activation in that region. Therefore, the risk is low, especially when the volume of the region is large enough to contain a lot of activation. In the same context, assessing neuronal death in the Hippocampus due to excitotoxicity in chronic epilepsy requires precise counting and, most importantly localization of neurons. 

The detection, localization and tracking of engrams, which are neurons collectively encoding stable representation of memory and have undergone enduring physical or chemical changes~\cite{tonegawa2018role, josselyn2020memory}, needs precision in their localization. Based on the experimental evidence gathered so far, the engrams, specific to context and environment~\cite{de2019optogenetic,pignatelli2019engram}, are sparsely populated and may possess distinct morphological (dendritic and axonal arbor spreadout) and electrophysiological properties (such as, short-time high excitability)~\cite{pignatelli2019engram}. High false negative and false positive detection rates substantially alter our understanding of the working principles of engrams.

\section{Applications}
\label{app}
Two major applications that extensively leverage neuronal cell counting and localization are finding neuronal circuitry during seizures~\cite{dabrowska2019parallel} and the detection of engrams (or engram complexes)~\cite{tonegawa2018role} for 3D memory mapping. Besides, there are region-selective cell-specific counting and localization, which aids in understanding the function of the cell population. One such example is the neocortical and hippocampal gamma-band oscillations, orchestrated by the spontaneous interplay between pyramidal cells and interneurons~\cite{buzsaki2012mechanisms}, during memory consolidation. 

During seizure, specific brain regions (striatum, thalamus, hippocampus etc.) and specific neuronal population (parvalbumin and somatostatin interneurons, dopaminergic neurons, GABAergic neurons etc.) are of prime interests. The false negative rate of detecting those cells should be significantly low in these scenarios. The size of the region containing specific neuronal population poses challenges and limits to obtain low false negative and false positive rates. For example, brain regions, such as dentate gyrus, ventrolateral thalamus and substantia nigra reticulata have relatively smaller volumes with high neuronal density and diversity. Localization in such small-volume structures with densely packed neurons needs precision of detection algorithms. 

Along with behavioral scores,  quantitative estimates of the number of specific cell types in a region, activated under control and pathogenic conditions, is an indicator of whether the region can serve as a suitable pharmacological target by optogenetic or chemogenetic inhibition or excitation, or a target for surgical ablation (such as, thalamotomy, pallidotomy~\cite{de2019pallidal}).   
For targeted therapeutic development, the variations in the cell counts (activated, degenerating or deceased) as functions of doses and concentration of a candidate drug, administered to the animals, are necessary for evaluating a drug's efficacy.

\section{Methods}
\noindent\textbf{Network Architecture}\\
Cellcounter is an encoder-decoder type deep neural network, consisting of a sequence of convolutional layers and an attention module. The encoding stage comprises $7$ convolutional layers, computing deep features at multiple, successive scales (sizes). The number of filter channels in consecutive convolution layers is increased ($32\longrightarrow 128\longrightarrow 32\longrightarrow 512\longrightarrow 32\longrightarrow 1024$), implying feature extraction at deeper levels, with the reduction in the size of the filtered output. The reduction in size was performed during convolution by setting the stride length of each filter channel as $2$. Then, each filtered output is batch-normalized and passed through a ReLU layer. 
Batch normalization reduces the effect of covariance shift triggered by different scales of convolutional filter-weights. 
To minimize the number of parameters, contraction layers are used. For example, filters in the successive layers in  $...32\longrightarrow128\longrightarrow32...$ has a contraction layer right after the $128$ filter convolution layer. The merge layer (Fig.~\ref{fig1: workflow}, top) concatenates the current filtered output with the one (downsampled version) at the previous layer along the dimension of filter length. The downsampling is performed by using a max pool layer. Each merge layer assembles information at two consecutive scales and the following convolution layer processes such combined-scale information.   

The construction of output map using information from a fine-scale structural scale at the last encoding stage is done in the decoding stage. The decoding stage consists of $6$ convolutional layers with upsampling layers between consecutive convolutional layers, producing the output per-pixel regression map in steps. Above a certain threshold, the regression map identifies the neuron cell bodies. Upsampling module includes nearest neighborhood based pixel interpolation. The number of parameters and the output dimension of each convolution layer are presented in Table~\ref{cellarch}. 

The attention module is based on the the principle of non-local weighted based averaging of pixels.
By non-local mean, attention module outputs a map that identifies salient regions. 
Let the input layer to the attention module be $x\in \mathcal{R}^{H\times W\times F}$, where $H$, $W$ and $F$ denote the height, width and the number of filters respectively. The attention of $x$ is defined as $x\longrightarrow softmax(A(x)B(x)^T)C(x)$. Here, $A$, $B$ and $C$ are linear embeddings. In our design, we take identical embeddings for $A$, $B$ and $C$, and $A(x)$ simply reshapes the layer as $x^{'}\in\mathcal{R}^{HW\times D}$ without using any bottleneck of the number of filters $D$. The attention module is placed before the final layer of the encoding stage so that the decoding stage would follow the maps containing salient regions.   

Cellcounter is not as deep as other state-of-the-art neural networks, such as U-Net ($26$ hidden layers), making it amenable for faster training with a moderate data size. There are a total of $1,431,255$ trainable parameters, with the forward/backward pass size of 24mb. There are no non-trainable parameters. The pretrained network is dependent on the patch size owing to the presence of the attention module. We trained Cellcounter with patches of size $64\times 64$. Users can also train the network with patches of a different size. While counting on an unknown test data, the test image is partitioned into a running sequence of non-overlapped patches of size $64\times 64$ and Cellcounter produces output maps for each of them, followed by image stitching to obtain the full output map. 
\begin{table*}[ht]
\vspace{-.1cm}	
	\caption{\small Cellcounter architecture details. Total parameters = 1431255, trainable parameters = 1431255. For a patch of size $64\times 64$, the forward/backward load = 23.64 MB, parameter load = 5.46 MB. }
	\label{cellarch}
	\centering
	\begin{tabular}{|l|c|c|}
	\hline
		Layer & Shape  & Parameters \\\hline 
		Conv2d-1   &$[-1\times 32\times 64\times 64]$  &320    \\\hline
		Conv2d-3 &$[-1\times 128\times 64\times 64]$   &36992    \\\hline
		Conv2d-5   &$[-1\times 32\times 32\times 32]$  &36896    \\\hline
		Conv2d-8  &$[-1\times 512\times 32\times 32]$  &295424    \\\hline
		Conv2d-10  &$[-1\times 32\times 16\times 16]$  &147488    \\\hline
		Conv2d-13 &$[-1\times 1024\times 16\times 16]$ &590848    \\\hline
		Attention && \\\hline
		Conv2d-15  &$[-1\times 32\times 8\times 8]$    &294944     \\\hline
		Conv2d-18  &$[-1\times 32\times 8\times 8]$   &18464     \\\hline
		Conv2d-20  &$[-1\times 16\times 16\times 16]$  &4624     \\\hline 
		Conv2d-22 &$[-1\times 8\times 32\times 32]$   &1160     \\\hline 
		Conv2d-24 &$[-1\times 4\times 64\times 64]$ &292  \\\hline 
		Conv2d-26 &$[-1\times 2\times 64\times 64]$ &74   \\\hline 
		Conv2d-28  &$[-1\times 1\times 64\times 64]$  &19     \\\hline 
	\end{tabular}
	\vspace{-.3cm}
\end{table*}

\vspace{2cm}
\noindent\textit{Data composition and augmentation}\\

All Training, validation and learning data have three components - image patches, repel-coded labeled images, text files containing the 2D locations of neurons, which are called through a dictionary format. Image patches are normalized with intensity values between $0$ and $1$, by using
\begin{eqnarray}
I_{normalized} = \frac{I - min(I)}{max(I)-min(I)}; max(I)>min(I).
\end{eqnarray}
For a $.nd2$ multi-channel, multidimensional image file with metadata, the images are first read using ImageJ (\textit{https://imagej.net/Fiji}) and the associated $Z$ stack with individual channel images are exported with $.tiff$ extension to a folder. In many instances, a $.tiff$ image file appears blank even though it contains considerable information about the slice. This happens because $.tiff$ supports $16$-bit data and it requires the conversion to $8$-bit ($uint8$). We provided a MATLAB code to accomplish the conversion.

We process Dapi and tdTomato channels to obtain Mixed data. The steps are outlined below.\\
\begin{algorithm}[H]
\SetAlgoLined
\SetKwInOut{Input}{input}\SetKwInOut{Output}{output}
\KwData{$I_{dapi}, I_{tdT}$}
 \Input{tdT-thresh, tdT-min-level, gray-gap, pix-thresh}
\Output{OutImage}
 locT = find($I_{tdT} > $tdT-thresh); locP = find($I_{tdT} < $tdT-thresh)\;
 OutImage(locT) = $I_{tdT}$(locT)\;
 conncomp = connected-component(OutImage)\;
 locNoise = find(conncomp$<$pix-thresh)\;
 locP = ADD(locP,locNoise); locT = REMOVE(locT,locNoise); tmp = $I_{tdT}$(locT)\;
 OutImage(locT) = $\frac{tmp - \text{tdT-thresh}}{max(tmp) - \text{tdT-thresh}}(255-\text{tdT-min-level}) )+ \text{tdT-min-level}$\;
 $I_{dapi} = \text{anisotropic-diff}\big(\frac{I_{dapi}}{255}*(\text{tdT-min-level}-\text{gray-gap})\big)$\;
 OutImage(locP) = $I_{dapi}$(locP);
 \caption{Fuse multichannel data: Mixed data}
\end{algorithm}
\vspace{.5cm}
Data augmentation is instrumental to training data-hungry deep network with only a few manually annotated image patches. However, biological constraints do not permit the use of all the conventional means to augment data for neurons. The morphology of a neuron primarily depends on its regional location, function and cell types. Data augmentation techniques, such as scaling and resizing of image patches induce unwanted distortions in neuronal shape, whereas rotation and translation preserve the neuronal morphology. We applied a fixed set of rotations on each image patch to carry out data augmentation and the procedure was executed \textit{on-the-fly} to alleviate the running storage load of the model during training.

\vspace{2cm}
\noindent\textit{Cellbody detection}\\
For each image input, Cellcounter yields a regression map as the output, which is then subjected to a routine that finds peak local maximum response.  The module is provided in skimage in scipy (\textit{https://scikit-image.org/docs/dev/api/skimage} ). 
It detects local maxima in a region of $(2 * min-distance + 1)$ so that peaks are separated by at least $min-distance$. We set $min-distance=2$, implying that two neurons whose cellbodies are $1$ pixel apart are not detectable by Cellcounter. This is a hard restriction and users can relax it to more than $2$ pixels in order to obtain improved results. the reason for our selection of $min-distance=2$ is that we observed multiple groups of neurons are overlapped with their cell bodies appearing in the close proximity of one another in regions, such as thalamus, entorhinal cortex and amygdala in coronal slices. 

Manual errors are unavoidable during annotation, where types of errors typically include labels away from the cell body and multiple labels on the same cell. During Phase 2, prior to updating the labels of newly detected cell locations, a cell-location-refinement routine is run to repeatedly adjust the labels of manually annotated cells in order to minimize the false negatives. In this routine, if the detected location of a neuron is in the proximity of its manual annotation, the manual annotation is updated in a search window. 

Cell center coding plays a crucial role in Cellcounter's performance. The center coding provides the output labels during training. We employed repel coding strategy, where the kernelized labels of two annotated neurons are well-separated if they are overlapped or extremely close to each other in an image. In this way, this coding strategy ensures good reversibility. The repel coding scheme is 
\begin{eqnarray}
D_{ij} &=& d^{1}_{ij}\times\Big(1 + \frac{d^{1}_{ij}}{d^{2}_{ij}}\Big)^2; d^{2}_{ij}\neq 0,\nonumber\\
C_{ij} &=& \begin{cases}
\frac{1}{1+\kappa D_{ij}};~if~D_{ij}<r,\\
0; \quad otherwise.
\end{cases}
\end{eqnarray}
Here, $(i,j)$ is the coordinate of a pixel. $d^{1}_{ij}$ and $d^{2}_{ij}$ are the distances between $(i,j)$ to its nearest and second nearest cell centers. $\kappa$ is a constant. $\kappa$ and $r$ control the width of the kernel. We set $\kappa = 0.8$ and $r = 9$. 

\vspace{1cm}
\noindent\textit{Training} \\
Cellcounter is trained on an nVidia GeForce GTX 1080 Ti with GDDR5X RAM, and scripted in python (version 3.7) using the Pytorch platform (version $10.2$) and cuda (version $10.2$). However, one can apply Cellcounter on test images without using cuda if the Pytorch (cpu version) is installed.  
The images are read as well saved using OpenCV (version 3.4.2). Images that are used in training, validation and testing are saved in $.tiff$ format. We performed maximum intensity projection (MIP) routine on the Z-stack to obtain a single image per $.nd2$ file. Later, we proceeded to sampling neurons from the MIP image. 

The network parameters are initialized with normal distribution, also known as He initialization. The resulting weights have values sampled from a normal distribution with zero mean and variance of $\frac{2}{fan-in}$. 
We used the rotation set $\Big[$ 233, 258,  62, 196, 106, 259, 242, 355, 320, 222, 313, 247, 315, 102,  24, 140, 66,  72, 76, 138, 326, 351, 330,   5, 356, 225, 311, 234,  78, 280,  89, 309, 279,  40, 292, 227, 111, 116, 268, 262, 334, 177,  85, 37, 187, 127, 142, 275, 354, 218$\Big]$ (in degrees). For each image, subjected to a rotation, the repel-coded label map is created $on-the-fly$.
In each epoch during training, all the data are first rotated using each rotation and divided into a number of batches to channel into the Cellcounter and this procedure is executed for all the rotations. In each epoch, each rotation is called in a random fashion to nullify any bias in the selection of an entry from the rotation set. Following this procedure, Cellcounter is additionally trained with the original data. The training loss that is incurred by the original data and augmented data is computed as the mean square error between the output maps and the corresponding repel-coded label images. We set adaptive moment estimation (Adam) with a user-specified learning rate as the optimization method for Cellcounter. After each epoch, the evaluation metric scores are displayed along with the training and validation losses. 

The training schedule (both in Phase 1 and 2) is divided into several sessions and each session has a fixed number of epochs and a learning rate, the selection of which is guided by the visual inspection of the metric scores in its previous session. In each session during training, the configuration that yields high recall on the validation set and relatively smaller difference between the precision and expected precision scores is saved and further utilized as the pre-trained network for the next session. There is an additional computational cost that is involved in Phase 2 - updating the labels. As it is cumbersome to update after every epoch, Cellcounter offers flexibility to adjust the start, interval and end of the epochs at which the update process is executed.

\vspace{1cm}
\noindent\textit{Evaluation metrics} \\
To systematically quantify the progression of self-learning as well as the performance on the validation set, a suite of measures are used - relative precision, expected precision, relative recall, precision, recall and F1 scores. 
Relative precision, expected precision and relative recall are used in the context of self-learning, whereas the rest are used to assess Cellcounter on validation and test datasets. Precision, recall and F1 scores are defined as follows.
\begin{eqnarray}
recall &=& \frac{\text{True positive}}{\text{True positive}+\text{False negative}}\nonumber\\
precision &=& \frac{\text{True positive}}{\text{True positive}+\text{False positive}}\nonumber\\
F1 &=& 2\Big(\frac{\text{precision*recall}}{\text{precision}+\text{recall}}\Big)\nonumber
\end{eqnarray}
$\text{True positive}$ or TP is defined as the number of neurons which are also detected by Cellcounter. $\text{False negative}$ or FN refers to the number of neurons which are undetected by Cellcounter. $\text{False positive}$ or FP denotes the number of locations which are falsely identified as neurons by Cellcounter. An efficient localization algorithm is expected to reduce the $\text{False negative}$ and $\text{False positive}$ scores. 

The term, `relative' in measures during self-learning stems from the incomplete sampling of neurons per patch and it refers to the computation the standard measures - precision and recall - with respect to the manually annotated labels. 

\begin{eqnarray}
relative~recall &=& \frac{\text{TP w.r.t. annotation}}{\text{TP w.r.t. annotation}+\text{FN w.r.t. annotation}}\nonumber\\
expected~precision &=& \frac{\text{TP w.r.t. annotation}}{\text{TP w.r.t. annotation}+\text{FP w.r.t. annotation}}\nonumber\\
relative~precision &=& \frac{\text{TP w.r.t. annotation}}{\text{TP w.r.t. annotation}+\text{FP by Cellcounter}} 
\end{eqnarray}

If a small number of neurons in an image is manually labeled, relative recall score would largely undermine the rest, which are unlabeled. With respect to the annotations, unlabeled neurons would be considered as FP. Therefore, expected precision score accurately reflects on how many unlabeled neurons Cellcounter needs to detect, whereas relative precision score indicates how many of them are detected by Cellcounter so far.

\vspace{1cm}
\noindent\textit{Testing} \\
Each test image is partitioned into patches of size $64\times 64$. Cellcounter labels the neurons in each patch. After finishing the labeling step, to retrieve the original image all the patches are assembled in the order they are partitioned . It is important to note that Cellcounter works only on gray valued images and for each gray valued image, it produces a color image by replicating the gray channel three times for R,G and B, with the red dots as the neuron cellbodies. The corresponding `Z' depth is saved in a separate .mat file. The test execution time is given in Table~\ref{testtime}. 

\begin{table*}[ht]
\vspace{-.1cm}	
	\caption{\small Execution time of Cellcounter when applied on an image of dimension $7460\times 5301$.}
	\label{testtime}
	\centering
	\begin{tabular}{|l|c|c|}
	\hline
		Configuration  & OS & time (in min) \\\hline 
		nVidia GeForce GTX 1080 Ti,& Linux (Ubuntu 18.04.5 LTS)  & 3.6 mins \\
		 Cuda 10.2, Driver 440.10  & Kernel = 5.4.0-42-generic&\\\hline
		\text{Intel(R) Core(TM) i7-8700K} &Linux (Ubuntu 18.04.5 LTS) & 11.04 min \\
		 CPU at 3.70GH &Kernel = 5.4.0-42-generic&\\\hline
		Intel(R) Xeon(R) CPU E5-1620  &Windows 10 pro, 64 bit & 14.7 min\\
		v4@3.50 GHz & & \\\hline
	\end{tabular}
	\vspace{-.3cm}
\end{table*}

\section{Results and discussion}
\subsection{Dataset description}
In this work, we demonstrate the efficacy of our model on primarily four datasets, shown in Fig.~\ref{fig2: Data}(F), which are primarily datasets containing tdTomato-expressed neurons in Trap2 mice. The result on NeuN-stained data is presented in \textbf{Supplementary Fig. 2}. 
We also notice that, in a number of experiments, the registration of a brain is prerequisite for reconstructing the mouse brain in 3D, prior to locating the neuron cells. Stainings, such as Dapi and NeuN preserve structural cues (edge, boundary of a brain region etc.) that aid in improving the quality of registration. Aiming at such applications, we have also curated a dataset, called `Mixed data', by disseminating the structural information of brain regions by combining two image channels (Dapi and tdTomato)followed by registration. 
In Fig.~\ref{fig2: Data}, images (B)(e-f), (E)(e-f) were sampled patches from the Mixed data. Although the fusion of different modalities introduces artifacts and variations (blur, gray value saturation, axon pruning), we encounter that the performance of Cellcounter is marginally affected when it is properly trained. In images, neurons appear to have variations in a number of aspects including morphology (B(a-f)), number of pixels delineating a neuron, pixel saturation level (B(b-c), C(a-f)) and clustering pattern (B(d-f)) (see Fig.~\ref{fig2: Data} for details).

\begin{figure}[!ht]
\vspace{-.2cm}	
	\centering
	\includegraphics[width=14cm, height=17.5cm]{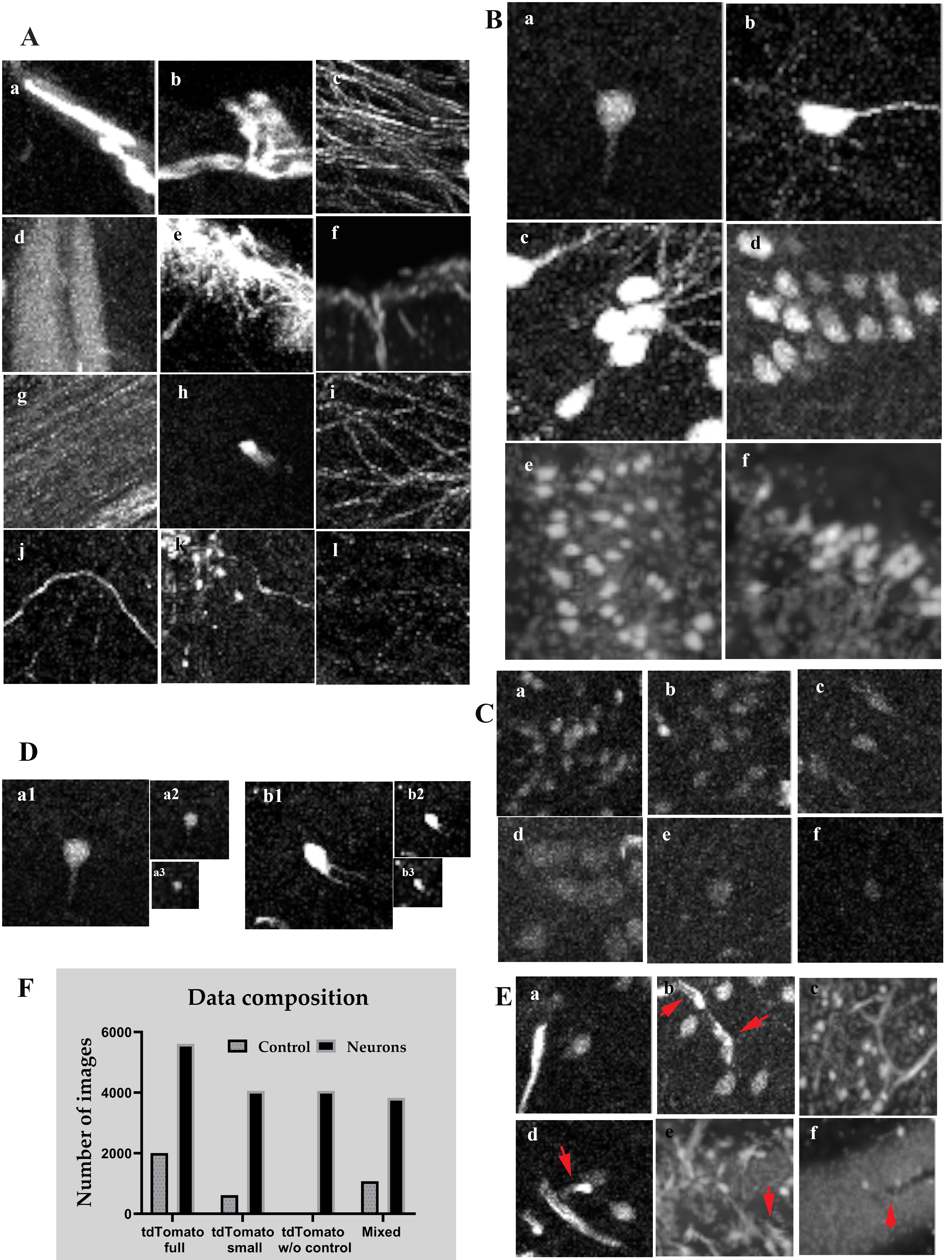}
	\caption{}
	\label{fig2: Data}
	\vspace{-.4cm}
\end{figure} 
\begin{figure}[!ht]
  \contcaption{\small (\textbf{A})(a-l) \textit{A collection of $12$ images ($64 \times 64$) depicting image background}. These are used as controls (axonal fibers, splintered parts, blotch etc.). (\textbf{B})(a-f) \textit{$6$ images ($64\times 64$) displaying variations in neuronal morphology and spatial allocation with orientation} - (a) pixilated (a heterogeneous mixture of gray values of contiguous pixels), (b) saturated in gray values, (c) cluster of saturated neurons, (d) cluster with varying intensities of axons, (e) blurred neurons with defined cell boundaries and (f) blurred and overlapped neurons. Note that (a-d) are sampled from tdTomato data, whereas (e-f) are from Mixed data . (\textbf{C})(a-f) \textit{A collection of low-intensity neurons}. These are ubiquitous in certain areas, such as thalamus in coronal slices.(\textbf{D}) \textit{Downsampling of images}. After applying successive downsampling (a1-a3 and b1-b3) on images, neurons are indistinguishable from control (such as, \textbf{A}(h)). (\textbf{E}) \textit{Spatial co-existence of controls with neurons.} The arrows in (b) point at where controls camouflage as neurons. Axonal arbors occlude neurons in (e).  The granule cell layer in the Dentate Gyrus (DG) imprints a bright, continuous background in (f).  (\textbf{F})\emph{Composition of datasets.}  The optimal result on the localization of tdTomato-expressing neurons, during self-learning on the learning set and generalization on test data, is obtained from the first dataset (tdTomato full), containing $2006$ controls without neurons and $5614$ images containing incompletely-labeled neurons. We have also investigated cellcounter on the Mixed data ($1077$ controls and $3826$ images with neurons) to study the effect of controls.    }% Continued caption
\end{figure}

\subsection{Self-learning capacity of Cellcounter}
\label{self-learning}
We provide partially labeled or annotated input images to Cellcounter, where `incompleteness' in labels implies that, a subset of neurons in each $64\times 64$ image patch is labeled. We define sampling density of a patch as $\frac{\text{number of neurons labeled}}{\text{number of neurons present}}$ and have plotted the histogram in Fig.~\ref{fig5: MoreResults} for tdTomato full dataset. It can be observed that image patches with relatively larger sampling density populate the dataset. In Fig.~\ref{fig3: Analysis}, the image patches in the first row, sampled from Mixed data, present three such examples, with the second row displaying the number of neurons that are initially annotated. Other neurons were progressively labeled by Cellcounter itself as the training continues.

To highlight the self-learning capacity in datasets, (for example, tdTomato data) we inspect a small subset ($89$ image patches from tdTomato full dataset), called `learning set', of training data over epochs. Ideally, all the training data at a training instance should be checked to verify such ability. However, it would essentially decelerate the training process. The learning set, which is a representative set of the large training set, is procured from different regions of various brain slices and it collectively encompasses enough variation to test self-learning. Simultaneously, we also examine the results on the validation dataset ($173$ image patches, tdTomato full dataset), which is separate from the training dataset, to assess how progressive labeling during training affects the generalization.

We divide the training schedule of Cellcounter in two phases - training with no update of labels (Phase 1) and training with update of labels (Phase 2). In each of the top four plots of Fig.~\ref{fig3: Analysis} the transition between Phase 1 and 2 is marked with an inverted `T' (thick black). In Phase 2, after an epoch, each training patch is evaluated by Cellcounter that has labeled a set of neurons and the corresponding label image is generated for further training in the following epoch. 
To quantitatively assess the self-learning ability, we define a set of metrics (see \textbf{Online Methods}) -  expected precision and relative precision and relative recall. To measure the generalization capacity of Cellcounter on the validation and test datasets, we follow the standard metrics - precision, recall and F1 score- for comparison. 

\begin{figure}[!ht]
\vspace{.2cm}	
	\centering
	\includegraphics[width=12cm, height=16.5cm]{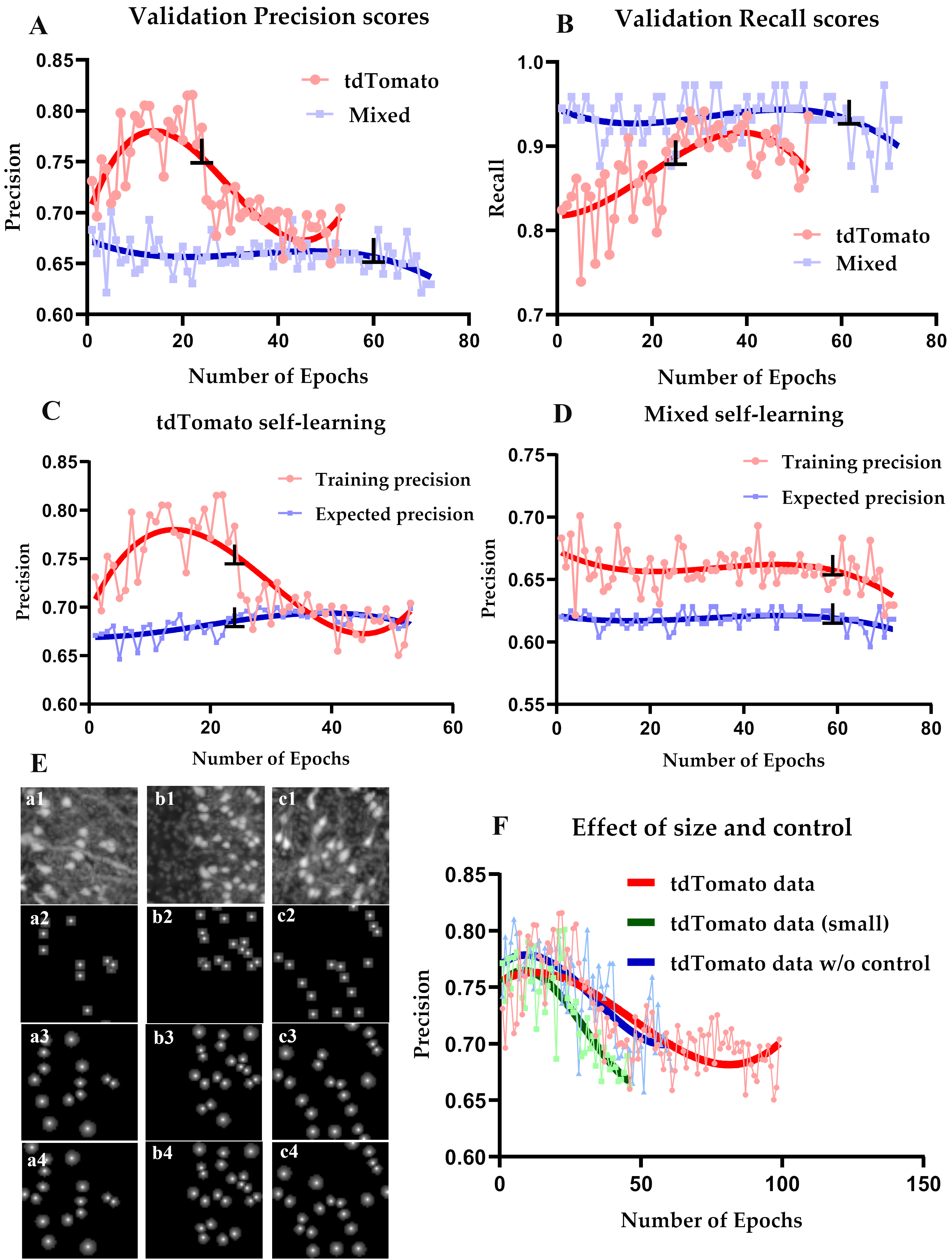}
	\caption{}
	\label{fig3: Analysis}
	\vspace{-.4cm}
\end{figure} 
\begin{figure}[!ht]
  \contcaption{\small Plots showing the quantitative analysis of the performance of Cellcounter during training, self-learning and validation. The precision scores are \textit{relative} to the manual labels. (A-B) Phase 1 (no change/update of training labels) and Phase 2 (update of training labels as predicted by Cellcounter itself) are marked by the inverted black `T' in the plots (tdTomato: $173$ $64 \times 64$ image patches, $387$ labeled and $169$ unlabeled neurons; Mixed: $103$ $64 \times 64$ image patches, $441$ labeled and $218$ unlabeled neurons). The expected precision values at recall score $1.0$ are $0.68$ and $0.67$ for tdTomato and Mixed datasets respectively. (C-D) explain the self-learning capacity of Cellcounter. The effect of data size and control are also tested on Cellcounter. Even with controls, it appears that smaller dataset leads to a substantial increase in the number of false positives, as evident from the plummet of precision (E), three images (a,b and c) at four different epochs are shown (Mixed data) to explain how Cellcounter predicts and relabels its training data progressively with epochs.      }% Continued caption
  \vspace{.1cm}
\end{figure}

The following example will clarify the need for the metrics to measure self-learning ability.
Let us assume that a patch contains $15$ neurons, $6$ of which are manually labeled and $9$ unlabeled. Relative to the existing labels in the patch, $6$ cells are neurons and other locations, detected as neurons by Cellcounter, would be treated as false positives.  
With this setting at hand, assume that Cellcounter has detected $8$ locations, $5$ of which belong to the labeled neurons and $3$ belong to the unlabeled ones. The relative recall and the relative precision scores would be $\frac{5}{6}= 0.83$ and $\frac{5}{5+3}=0.625$ respectively. 
Notice that the relative precision score severely underestimates the true number of unlabeled neurons. Expected precision ($\frac{5}{5+9} = 0.36$) incorporates the unlabeled population. The difference of $0.625-0.36 = 0.265$ indicates that we need to allow Cellcounter to generate false positives relative to the manual annotation, so that the false positives are \textit{actual} unlabeled neurons. In Phase 1 (no self update of labels), we rigorously trained Cellcounter to  achieve a high relative recall value ($\sim 90\%$). A high relative recall value  is a potential indicator of capturing salient representations of neurons by Cellcounter. Once this recall score is achieved, we proceed to Phase 2, where labels are updated over epochs.

 Fig.~\ref{fig3: Analysis}(C-D) describes the self-learning assessment of Cellcounter. For quantitative assessment, we have examined both the training phases on the learning dataset to see  whether there is a difference in the scores between Phase 1 and 2. The precision scores on the learning set turns out to be non-normal (Shapiro-Wilks test, $P<0.0068$ for tdTomato data and $P<0.0013$ for mixed data). For tdTomato data, Phase 1 and Phase 2 are found to be different (two sample KS test,$P<0.000013$) in terms of the difference between precision and expected precision scores. This is same for the Mixed data (two sample KS test, $P<0.035$). This result validates our claim that the difference indeed approaches narrower with the continuation of label update during training, which is indicative of self-learning. Visual verification of all the patches from the learning dataset in Fig.~\ref{fig3: Analysis}(E) suggests that there are no false positives and unlabeled neurons are progressively detected. 

\begin{figure}[!ht]
\vspace{-.2cm}	
	\centering
	\includegraphics[width=12cm, height=17.5cm]{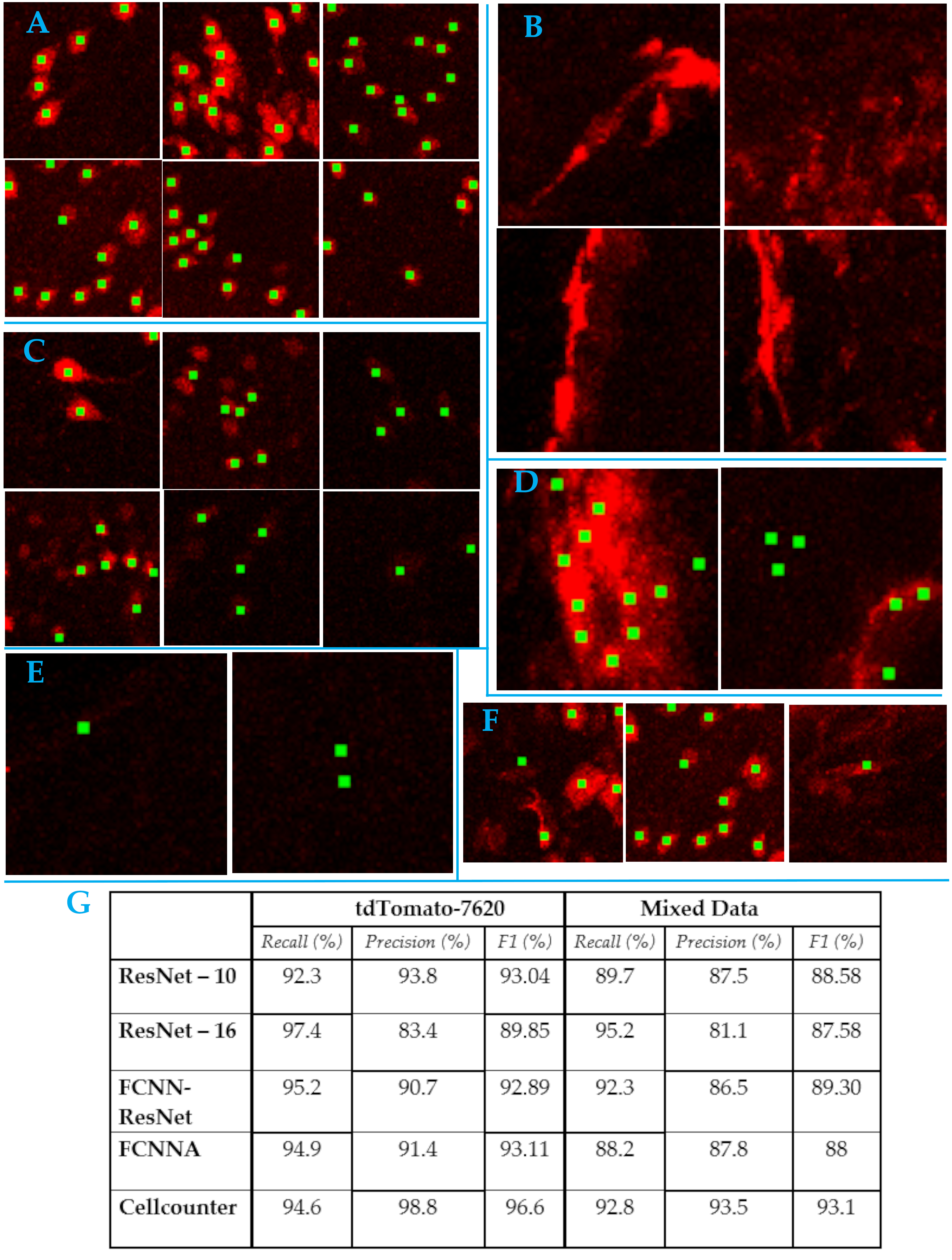}
	\caption{  }
	\label{fig2: Dataa}
	\vspace{-.4cm}
\end{figure} 
\begin{figure}[t]
  \contcaption{ \small (\textbf{A}) Figure shows six sample patches with different gray value composition and morphology of neurons, when Cellcounter is trained with sufficient control images. (\textbf{B}) Figure shows the result on control images, verifying the existence of almost no false positives on these non-neuronal structures. (\textbf{C}) Results on six image patches by Cellcounter, trained without control patches. Some of the detection results are noteworthy, such the last one in the bottom row, where neurons are indiscernible from the background. (\textbf{D}) Figure shows the performance of no-control-trained Cellcounter on the control patches. It clearly shows an enormous number of false positives. In addition, it frequently yields false positives on the background containing no visible structures, shown in (\textbf{E}). (\textbf{F}) Instance with unavoidable false positives and out-of-cellbody localization are presented, where Cellcounter is trained with enough control data. (\textbf{G}) Comparison of the performance against a set of state-of-the-art networks is given in a table. These networks are of similar size, but not as deep as the U-net.     
  }% Continued caption
\end{figure}

\subsection{Effect of data size and controls}
\label{ctrldata}
 Inclusion of such control images during training (Fig.~\ref{fig2: Data}(A)) has significantly improved the quality of prediction, as they suppressed and, in some cases, effectively eliminated the background information. The number of control images in each dataset is shown in Fig.~\ref{fig2: Data}(F). Fig.\ref{fig3: Analysis} presents the effect of data size and control images on the precision scores on the validation dataset. It can be observed that precision scores for all datasets, especially tdTomato with small data size have plunged over epochs. The recall scores are shown in Fig.~\ref{fig5: MoreResults}(A), where it can be seen that Cellcounter has performed better in smaller tdTomato data and data without control when compared to tdTomato full dataset. 
In short, in case of small size and without control datasets, the rise of recall as well as the drop in precision are accelerated, leading to the precipice of instability when Cellcounter starts yielding excessive false positive predictions without a sign of recovery.   

\begin{figure}[!ht]
\vspace{-.2cm}	
	\centering
	\includegraphics[width=12cm, height=16.5cm]{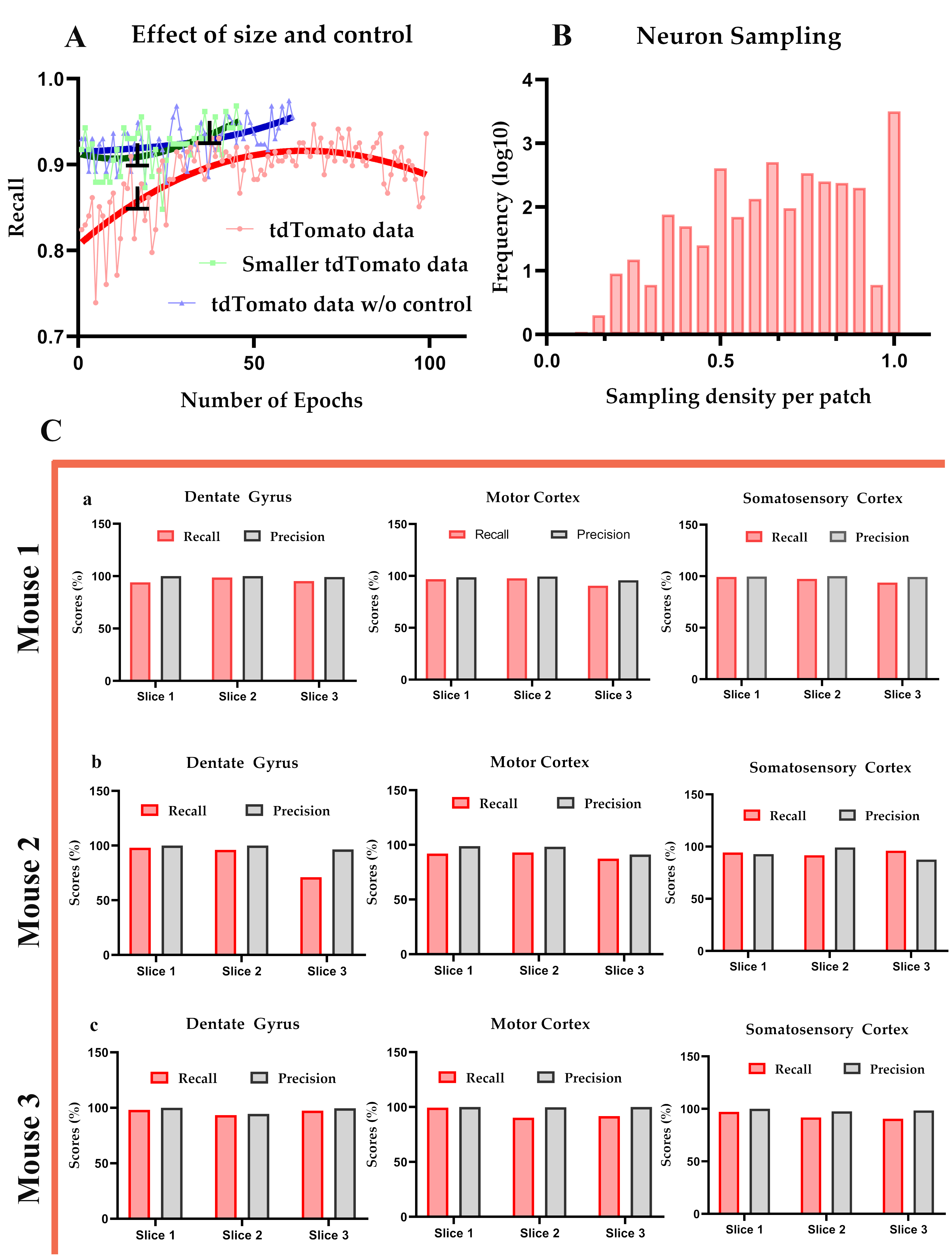}
	\caption{}
	\label{fig5: MoreResults}
	\vspace{-.4cm}
\end{figure} 
\begin{figure}[!ht]
  \contcaption{\small (A) exhibits the recall scores of Cellcounter during training on three different datasets. (B) represents the histogram of the neurons which are manually annotated. Note that the Y axis is in $log10$. The X axis represents the sampling density per patch ($\frac{\text{number of neurons labeled}}{\text{number of neurons present}}$).
 The sampling is expected to be independent of regional bias. For complete annotation, the frequency should be unity at sampling density $1$ and zeros at other values. It is evident that our annotation is incomplete.  (C) depicts the precision and recall scores of localization by Cellcounter on three slices taken from three TRAP2 mice. Cellcounter is tested on tdTomato datasets (Mouse 1 and 3) and a Mixed dataset (Mouse 2). The mixed dataset is more challenging because of the effects induced by the preprocessing steps (\textbf{Algorithm 1 - Online Methods}), registration (not shown) and Dapi channel information. Detection on tdTomato and Mixed datasets are shown in Fig.~\ref{fig4: Results} and \textbf{Supplementary Fig. 1}. We show the quantitative results on three brain regions - Dentate Gyrus, Somatosensory area and Motor cortex.  
  }% Continued caption
\end{figure} 

We have found that the metric scores are non-normal (Shapiro-Wilks test $P$ is $=0.0004$ for tdTomato data, $<0.0001$ for tdTomato small data and tdTomato without control). No difference of precision scores is found between tdTomato and without control (two sample KS test, $P = 0.1247$), whereas both of them are found different from small size tdTomato data (two sample KS test, $P=0.0002$). Next, we have inspected recall scores. Recall scores in case of tdTomato data is found different from others (two sample KS test, $P<0.0001$), whereas the test has not found a difference between the cases of tdTomato small and without control datasets (two sample KS test, $P=0.70$).  However, only the tdTomato data (large corpus; 7620 images) has restored the precision without moving past the expected precision. It ensures that Cellcounter is not prone to yield false positive labels. 

Effects of controls are shown in Fig.~\ref{fig2: Dataa}, where (A) and (C) present the results on patches taken from tdTomato datasets with and without control respectively. The drop in precision as a result of no control (Fig.~\ref{fig5: MoreResults}), is also evident in Fig.~\ref{fig2: Dataa}(C) because a number of neurons has not been detected by Cellcounter. (B) and (D) (Fig. \ref{fig2: Dataa}) present the results on images containing no neurons. It has been observed that Cellcounter trained without controls frequently spews false positives at dense structures, such as thick axonal fibers (Fig.~\ref{fig2: Dataa}(E)).

\begin{figure}[!ht]
\vspace{-.2cm}	
	\centering
	\includegraphics[width=12.5cm, height=17.5cm]{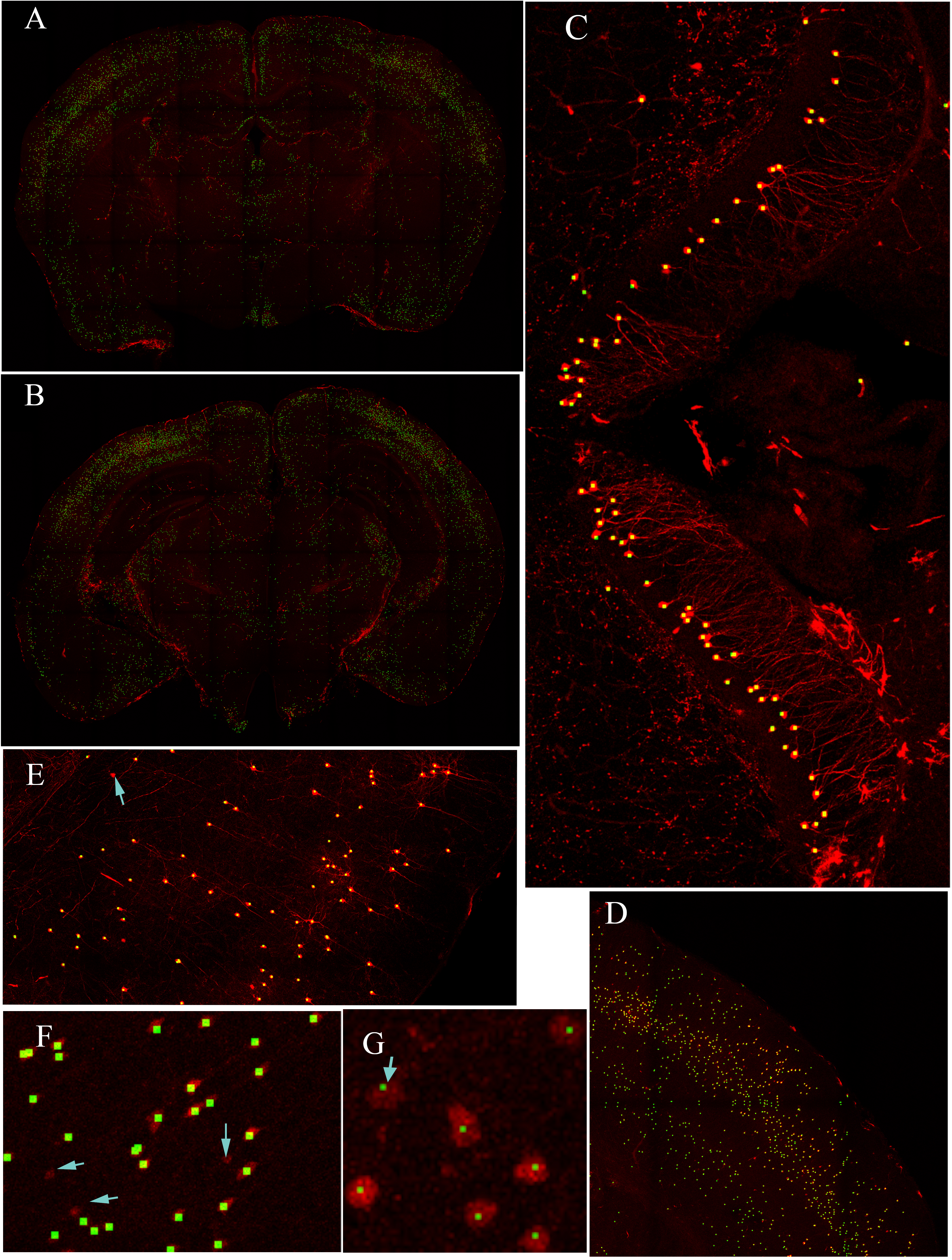}
	\caption{}
	\label{fig4: Results}
	\vspace{-.4cm}
\end{figure} 
\begin{figure}[!ht]
  \contcaption{\small Results on neuron localization by Cellcounter. Arrows in (E, F and G) are used to mark the false negatives. Each neuron location is shown with  $5\times 5$ square in green to enhance the visibility. However, the actual output labels each neuron with single green pixel, as shown in E and G. (A-B) exhibits the cell localization on two different slices with resolution $(8564\times 5788)$ and $(9482\times 7461)$ respectively.D shows the neurons, detected by Cellcounter, in somatosensory cortical layers.
  }% Continued caption
\end{figure}

\subsection{Detection and localization}
The results of neuron detection and localization by Cellcounter are presented in Fig.~\ref{fig4: Results} for tdTomato dataset and (\textbf{Supplementary Fig. 1} and \textbf{3}) for the Mixed and NeuN datasets respectively.

Labeled neurons (green square boxes) in two different MIP slices, containing hippocampal regions of two different TRAP2 mice are shown Fig.~\ref{fig4: Results}(A) and (B).
Fig.~\ref{fig4: Results}(C), obtained from a different TRAP2 mouse, displays the tdTomato-expressing neurons, labeled by Cellcounter, on the Dentate Gyrus (DG) in a $200\mu$ thick coronal MIP slice. Cellcounter perfectly labeled $67$ out of $69$ neurons and there are two false negatives and two false positive labels. One can see that neurons that are overlapped, contiguous without distinct cell boundaries in DG in the slice were also accurately labeled. The localization of neurons in part of the primary somatosensory area (barrel field, layer $1$ and $2$) is presented in Fig.~\ref{fig4: Results}(E), where there are two false negatives (one of them is shown with an arrow). 
Fig.~\ref{fig4: Results}(D) shows an example where there is no trace of false positives on the cell boundary. Fig.~\ref{fig4: Results}(F) captures a zoomed-in section in the primary motor area of a Trap2 mouse, where false negatives were shown with arrows in addition to the labeled neurons. Fig.~\ref{fig4: Results}(H) and Fig.~\ref{fig2: Data}(C-6) are alike in the sense that the neuron has low intensity, making it practically indistinguishable from the background. Cellcounter correctly localized the neuron with the position of the label having a little spatial offset from the cell center like the one pointed by an arrow in Fig.~\ref{fig4: Results}(G).

The accuracy of localization in case of Mixed dataset (\textbf{Supplementary Fig. 1}) is a little low compared to that of tdTomato datset, and this fact is also reflected in Fig.~\ref{fig5: MoreResults} (bottom) for Mouse 2. It is probably due to the adverse effects of the Dapi channel and the registration process. However, table G in Fig. \ref{fig2: Dataa} suggests that Cellcounter does not produce excessive false positive like other state-of-the-art deep networks.

We have evaluated the performance of Cellcounter in different coronal MIP slices taken from various brain regions. Each maximum intensity slice is obtained by projecting $20$ images of $10\mu$ optical thickness each onto a coronal plane. The projection also records the depth where maximum intensity at each pixel has occurred (Depth map; `z' coordinate). The $(x,y)$ coordinates of neuronal cellbodies are labeled by Cellcounter. We have inspected several regions and showed the quantification results on Dentate gyrus, motor cortex (primary and secondary) and primary somatosensory (trunk and barrel fields) regions in terms of precision and recall. 
The slices, with an example shown in Fig.~\ref{fig5: MoreResults}(bottom), altogether form the test dataset. Unlike the learning and the validation sets, all the neurons are manually labeled in each test slice. Therefore, we have followed the standard definition of precision and recall as $precision~=~\frac{true~positive}{true~positive+false~positive}$ and $recall~=~\frac{true~positive}{true~positive+false~negative}$. To perform consistent and unbiased comparison among the counts in regions over different mice, we have selected the coronal slices of identical coordinates (Atlas by Paxinos and Franklin) over all mice.

We have considered the data of tdTomato-expressing neurons in cases of Mouse 1 and 3, whereas we have examined slices of tdTomato fused with Dapi (Mixed data) in case of Mouse 2. A number of tests are performed to assert whether precision and recall scores are consistent over (1) slices and (2) mice for each region. When the counting and localization are performed in Dentate gyrus, the tests (two-way ANOVA) fail to reject the alternative hypothesis that there existed at least one slice in which the precision ($R>0.48$) and recall ($P=0.6$) are different at $5\%$ level of significance. Similar results were found for motor area ($R=0.13$ and $P=0.19$) and somatosensory cortex ($R=0.2946$ and $P=0.47$). In addition, We have not found significant differences in counts and localization over mice. The array of tables in Fig.~\ref{fig5: MoreResults} also shows that the average precision scores are also high, indicating effective suppression of false positive predictions.  

Comparison with the state-of-the-art networks is given in the table in (Fig.~\ref{fig2: Dataa}). It can be seen from the table that outperforms all the methods including Cellcounter in terms of Recall. However, it produces a lot of false positives. Cellcounter appears to greatly reduce the number of false positives.  

\vspace*{3cm}
\section{Conclusion}
Neuronal cell counting and localization have been long-standing challenges in computer vision and neuroscience because of enormous variation that neurons encompass in terms of imaging modality, anatomical features and processing protocols. It is one of the indispensable tools that are needed in several applications in neuroscience, especially where the formation of regional activation map or the assessment of cell death are required. Here, we presented Cellcounter, an easy-to-use, deeplearning based model to simultaneously detect, localize and record the coordinates of neurons. While doing so, we extensively procured representative image samples, containing neurons and backgrounds (controls) to robustly train Cellcounter. By construction, Cellcounter is at large data-hungry and it is extremely time-intensive to sample each neuron from a stack of slices. To ameliorate such manual load, we channeled image patches having incompletely and irregularly sampled neurons and augmented those images on-the-fly. Using this setting, we investigated the self-learning capacity and the generalization ability of Cellcounter. Armed with both of these features of Cellcounter, we analyzed its efficacy on various datasets. 

The necessity of self-learning stems from the training principle of a deep network and the incompleteness in our sampling. If the network possesses the ability to accurately predict a \emph{subset} of unlabeled neurons without yielding false positives after an epoch, then it is possible to complete the sampling of all the neurons in an image patch after gradually updating the labels over epochs. This was manifested when the precision touches the expected precision curve at a high recall ($\sim 0.94-0.97$) value. So, at first we introduced Phase 1 and carefully monitored if the network was giving false positives on the learning set. Then, the network was disambiguated in Phase 2 with regular update of training labels, with the expectation that the learning of unlabeled neurons was gradual in fashion. It can be hypothesized that after each update, the confidence of Cellcounter to label the remaining unlabeled neurons increases. 
Summarizing the above arguments, self-learning fulfilled the completeness in sampling and generalization was geared towards improving localization in test data, and both of them were performed simultaneously.

We catalogue a rich selection of background image patches containing noise, unwanted structures and artifacts. The effect of such control images on the performance during and test is shown in Fig.~\ref{fig3: Analysis} and~\ref{fig5: MoreResults}. In order to train Cellcounter on a new dataset, it is an advice to provide the model with an abundant supply of background images (see Fig.~\ref{fig2: Data}(A)). Incorporation of this collection of control images during training is a way to combat false positives. We have sorted out such control images for each modality of data that we used. However, excessive population of control images in a dataset can result in a spike of the number of false negatives. This is experienced when Cellcounter, trained using such setting, was used for prediction on the test data. The effect (occlusion, camouflage) of unwanted structures and artifacts appear due to the maximum intensity operation on `Z' stack. This adverse effect can be eliminated possibly to some extent by deep networks that are trained on 3D images. However, such 3D net will excessively increase the number of trainable parameters, ultimately demanding an enormous manually-annotated data. Therefore, efficient algorithms and intelligent network architectural design are needed to address localization in 3D.

\vspace{1cm}
\noindent\textit{Animals and Data}\\
We used brain slices imaged on a C2 Confocal (Nikon Instruments) from transgenic reporter mice (TRAP) expressing tdTomato under the control of cFos promoter to permanently label the activated neuronal population. Some of the brain slices involved DAPI and anti-NeuN antibody labelling to have different datasets that represent the base domain of all neuroscience lab data on neuron counting. For each modality, a sparse set of randomly-selected neurons were manually annotated from a set of brain slices taken from mice of both sexes. Centering each annotated neuron, a $64\times 64$ image patch was cropped to create the final dataset to be used in our model. All experiments and procedures were performed under the guidelines set by the University of Virginia Animal Care and Use Committee.

\vspace{1cm}
\noindent\textit{Code and Data availability}\\
The code and data are made publicly available in Github (https://github.com/50-Cent/Cellcounter), containing multiple modules including several stain-specific network configurations. We have also provided a user manual.

\section*{Author Contributions}
Tamal Batabyal developed the model. Aijaz Naik curated the data. Daniel Weller and Jaideep Kapur supervised the work. Tamal Batabyal, Aijaz Naik, Daniel Weller and Jaideep Kapur prepared the manuscript. 

\section*{Competing Financial Interests}
The authors declare no competing financial interests.

%\vspace{1cm}
%\noindent\textit{Data availability}\\

%\vspace{1cm}
%\noindent\textit{Acknowledgement}\\

%________________%
\section*{References}

%\bibliographystyle{unsrt}
%\bibliography{CellCountingBib}

\section*{Supplementary}
\begin{figure}[!ht]
\vspace{-.2cm}	
	\centering
	\includegraphics[width=12cm, height=16.5cm]{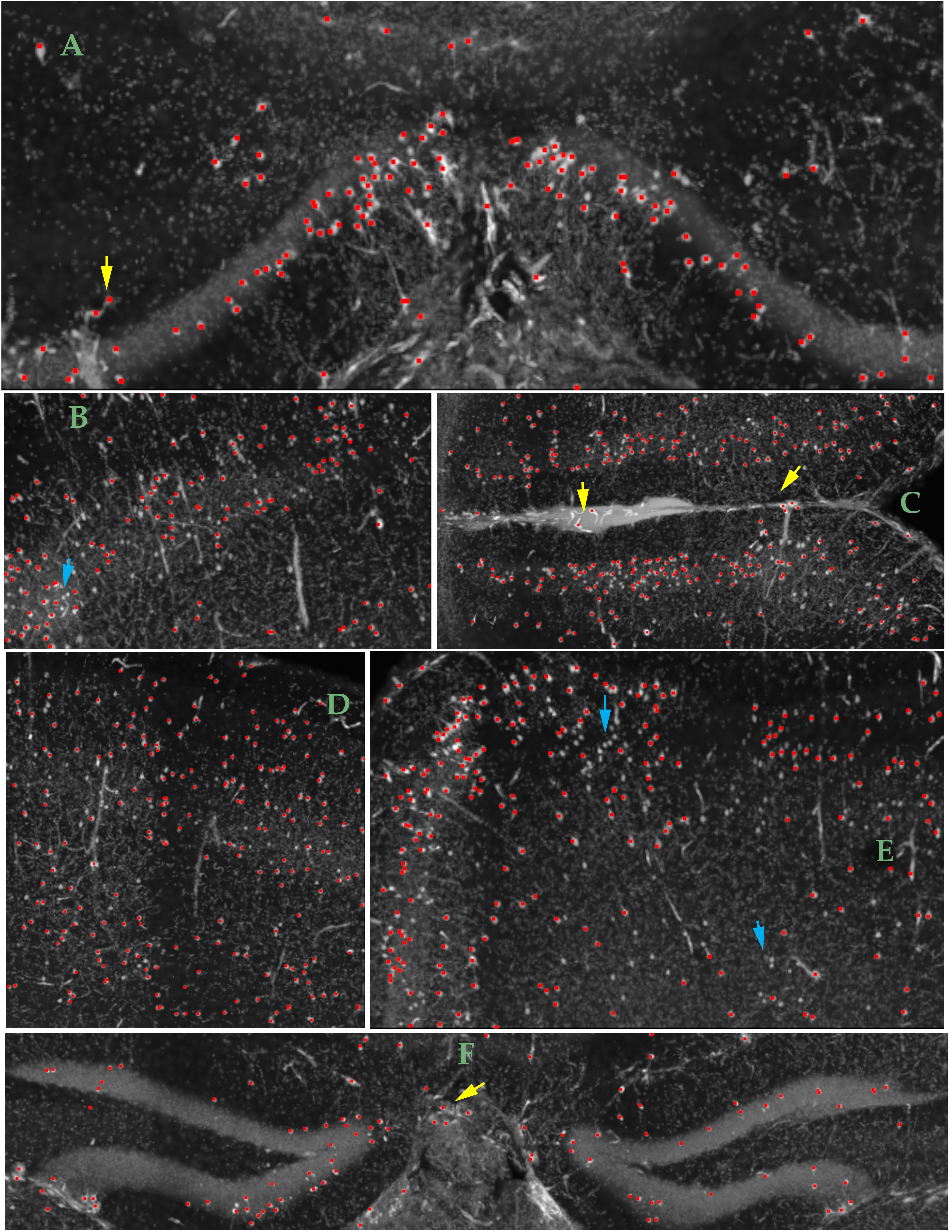}
	\caption{ }
	\label{fig2: Data}
	\vspace{-.4cm}
\end{figure} 
\begin{figure}[t]
  \contcaption{ This figure presents results on neuronal localization by Cellcounter on Mixed datasets. The regional structures as backgrounds are visible due to the presence of Dapi channel information. (\textbf{A}) and (\textbf{F}) display Dentate Gyrus with the neurons labeled by Cellcounter. (\textbf{C}) shows the result in the retrosplenial cortex, where (\textbf{B}), (\textbf{D}) and (\textbf{E}) are different cortical sections. Blue and yellow arrows point at a couple of false negatives and false positives.    
  }% Continued caption
\end{figure}

 \begin{figure}[ht]
\vspace{-.2cm}	
	\centering
	\includegraphics[width=12cm, height=16.5cm]{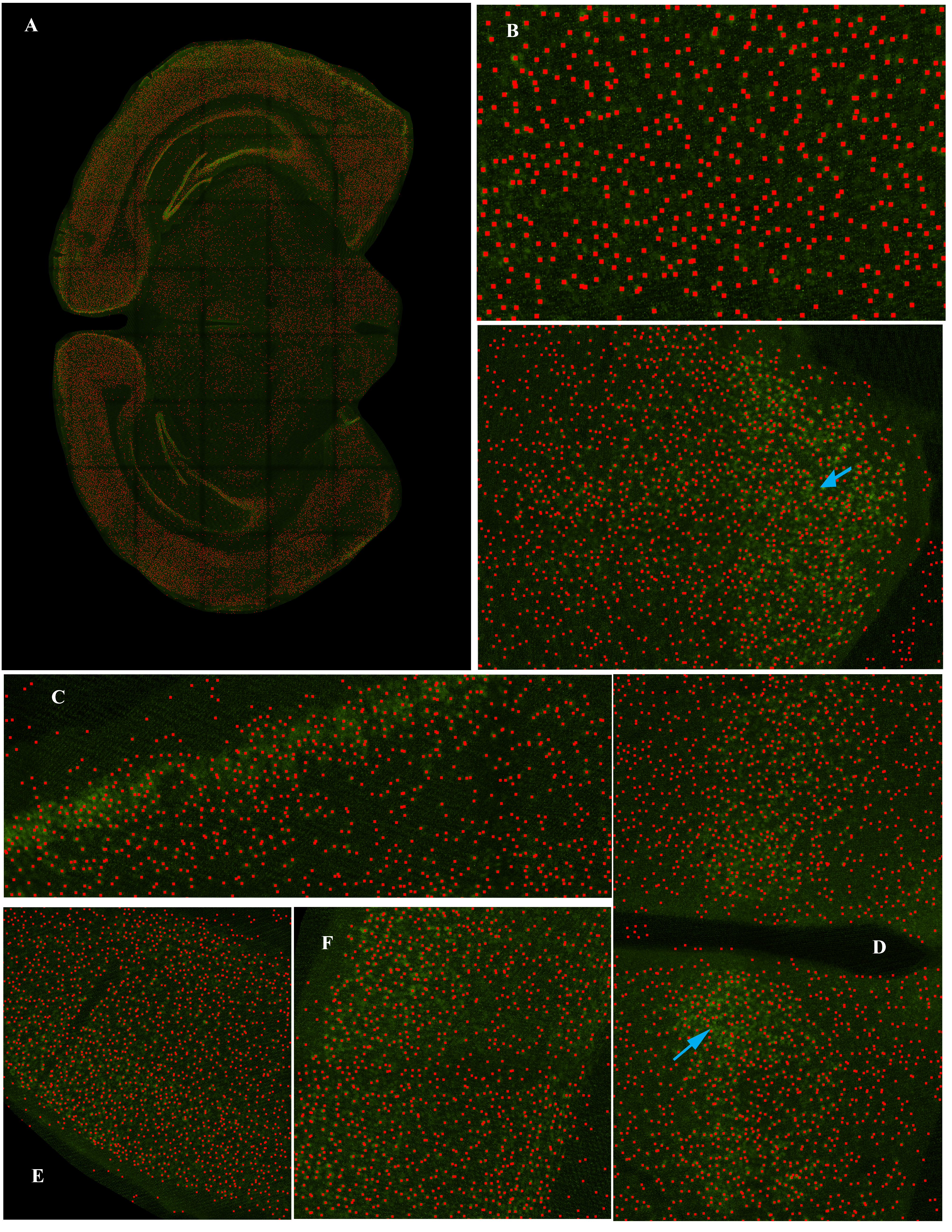}
	\caption{  }
	\label{fig2: Data2}
	\vspace{-.4cm}
\end{figure} 
\begin{figure}[t]
  \contcaption{ (\textbf{A}) This figure presents results on neuronal localization by Cellcounter on NeuN datasets. The red dots are the neuronal cell bodies.  
  (\textbf{B-G}) show the localization of neurons in the zoomed-in regions of different parts of a brain slice. The blue arrows point at a couple of false negatives With a good quality of NeuN staining, the recall and precision scores are 91.2\% and 95.7\%. 
  }% Continued caption
\end{figure}

\end{document}